\documentclass[structabstract]{aa}  
\usepackage{graphicx}
\usepackage{txfonts}
\usepackage{amssymb,amsmath,graphics,multirow,lscape}
\usepackage{natbib}
\bibpunct{(}{)}{;}{a}{}{,}
\usepackage{textcomp}
\usepackage{epsfig}
\usepackage{color}
\usepackage[hang]{subfigure}

\newcommand{\teff}{$T_\mathrm{eff}$}
\newcommand{\logg}{$\log g$}
\newcommand{\feh}{[Fe/H]}

\newcommand{\mic}{$\mu \mathrm m$}

\begin{document}

\title{Chemical Abundances  of M giants in the Galactic Center: \\ a Single Metal-Rich Population with Low [$\alpha$/Fe]\thanks{Based on observations collected at the European Southern Observatory, Chile, program number 089.B-0312(A)/VM/CRIRES and 089.B-0312(B)/VM/ISAAC}
}
\titlerunning{Chemical Abundances  of M giants in the Galactic Center}


\author{N. Ryde \inst{1}
\and   M. Schultheis \inst{2}
}
\institute{ Department of Astronomy and Theoretical Physics, Lund Observatory, Lund University, Box 43, 221 00, Lund, Sweden e-mail:ryde@astro.lu.se
 \and
	Observatoire de la C\^ote d'Azur, CNRS UMR 7293, BP4229, Laboratoire Lagrange, 06304 Nice Cedex 4, France 
 e-mail: mathias.schultheis@oca.eu
 }

 \date{Submitted 2013; accepted 2014}
 
\abstract
   {The formation and evolution of the Milky Way bulge is still largely an unanswered question. One of the most essential observables needed in its modelling are the metallicity distribution and the trends of the $\alpha$ elements as measured in stars. While Bulge regions beyond $R \gtrsim 50\,\mathrm{pc}$ of the centre has been targeted in several surveys, the central part has escaped detailed study due to the extreme extinction and crowding. The abundance gradients from the center are, however, of large diagnostic value.}
   {We aim at investigating the Galactic Centre environment by probing M giants in the field, avoiding supergiants and cluster members. }
   {For 9 field M-giants in the Galactic Centre region, we have obtained high- and low-resolution spectra observed simultaneously with CRIRES and ISAAC on UT1 and UT3 of the VLT. The low-resolution spectra provide a means of determining the effective temperatures, and the high-resolution spectra provide detailed abundances of Fe, Mg, Si, and Ca.}
   {We find a metal-rich population at [Fe/H]$=+0.11\pm0.15$ and a lack of the metal-poor population, found further out in the Bulge, corroborating earlier studies. Our [$\alpha$/Fe] element trends, however, show low values, following the outer Bulge trends. A possible exception of the [Ca/Fe] trend is found and needs further investigation.}
   {The results of the analysed field M-giants in the Galactic Centre region, excludes a scenario with rapid formation, in which SNIIe played a dominated role in the chemical enrichment of the gas. The metal-rich metallicities together with low $\alpha$-enhancement seems to indicate  a bar-like population perhaps related to the nuclear bar. }


\maketitle

\titlerunning{Low resolution K-band spectra}
\authorrunning{Schultheis \& Ryde}

\section{Introduction}
The formation of the Milky Way bulge remains a puzzle. Abundance ratios of key elements as measured in bulge stars of different populations and at different locations, can set empirical constraints on the formation and evolution of the Bulge \citep[see e.g.][]{silk:93,matteucci:12}. While more and
more detailed\footnote{at spectral resolutions of $R=\lambda/\Delta \lambda \gtrsim 50000$}  elemental abundance determinations  of stars in the intermediate and outer Bulge (such as
Baade's window)  have emerged, the study of detailed
abundances in Bulge regions within projected distances of $\rm R \lesssim 50\,pc$ of the Galactic Centre has until recently been prevented due to regions of  high extinction \citep[see e.g.][]{schultheis2009} and the lack of sensitive  high-resolution IR spectrographs. 
Extinction of 10-30 magnitudes in the V band can, however, be overcome by investigating stars in the K band, where 
the extinction is a factor of 10 lower \citep{cardelli}.

The abundance gradient with latitude is an important diagnostic of the properties and formation history of the Bulge \citep{grieco:12}. While the outer bulge shows evidence for a metallicity gradient \citep[see e.g.][]{hill2011}, be it due to a true gradient in a population or the relative change in  proportions of two or more populations, only a few studies have investigated abundance gradients in the inner bulge. \citet{rich2007} studied detailed abundances of M-giants in a field at $(l,b) = (0^{\circ},-1^{\circ})$, while \citet{rich:12} studied  fields located at lower latitudes: $(l,b) = (0^{\circ},-1.75^{\circ})$ and $(l,b) = (1^{\circ},-2.65^{\circ})$. They find an iron abundances of [Fe/H]=$-0.2\pm0.1$ for all three fields,  indicating a lack of any major
vertical abundance-gradient. A detailed abundance study in the Galactic Center (GC) region itself is missing. 
 Until now, most of the chemical abundance studies were limited to  luminous supergiants making the
abundance analysis difficult due to their complex and extended stellar atmospheres.
\citet{carr2000}, \citet{ramirez2000}, and \citet{davies2009} analysed high-resolution spectra  of supergiant stars in the GC region,  indicating metallicities of nearly solar metallicity. A similar conclusion was obtained by \citet{najarro2009} analysing two luminous blue variables. \citet{cunha2007} derived abundances of a few M giants 
in the Central Cluster and
found a metallicity of [Fe/H]$=+0.14\pm 0.06$, together with enhanced [O/Fe] and [Ca/Fe] abundances of $+0.2$ and $+0.3$, respectively, indicating that SN II played a dominant role  in the chemical enrichment of the gas for this cluster. 

 Here, we present an analysis, based on near-IR, high-resolution spectra, of 9 M-giant stars located in the GC, only $2.5-5.5\,$arcminutes  North of the very center of the Milky Way, which corresponds to a projected galactocentric distance of $5-10\,$\,pc. Our goal is to obtain the metallicity distribution and the corresponding $\alpha$-element abundances (Mg, Si, Ca) of GC stars. 


\begin{table*}
\caption{Basic data for the observed stars and their stellar parameters.\label{starsall}}
\begin{tabular}{l  c c c  c c c c c c c c}
\hline
\hline
\multicolumn{1}{l}{Star} & RA (J2000) & Dec (J2000) & $K^{a}$ & $K_0$ & exp. time &\teff & \logg & \feh &  $\xi_\mathrm{macro}$  &  $v_\mathrm{rad}$  & SNR   \\
         &    (h:m:s)    & (d:am:as)   &     &    &  [s] & [K] & (cgs) &   & km\,s$^{-1}$ &   km\,s$^{-1}$  & pixel$^{-1}$ \\         
\hline
\multicolumn{4}{l}{Galactic centre stars}\\
\hline
GC1   & 17:45:35.43 &  -28:57:19.28  & 11.90 & 9.40 & 3000 & 3668 & 1.37 & $+$0.19   & 6.6  &   $+$45.2 & 55   \\
GC20  & 17:45:34.95 &  -28:55:20.17  & 11.68 & 9.37 & 3600 & 3683 & 0.76 & $+$0.18   & 8.1  & $-$8.4   & 50     \\
GC22  & 17:45:42.41 &  -28:55:52.99  & 11.54 & 9.04& 3600 & 3618 & 0.70 & $+$0.02   & 6.0  & $-$123.5 & 90     \\
GC25  & 17:45:36.34 &  -28:54:50.41  & 11.60 & 9.10 & 2400 & 3340 & 0.77 & $+$0.08   & 7.6  & $-$9.7   & 50      \\
GC27  & 17:45:36.72 &  -28:54:52.37  & 11.64 & 9.14 & 3600 & 3404 & 1.16 & $+$0.31   & 8.1  & $-$210.8 & 50        \\
GC28  & 17:45:38.13 &  -28:54:58.32  & 11.67 & 9.17 & 3000 & 3773 & 1.24 & $-$0.13   & 5.7  &  $+$145.0 & 50        \\
GC29  & 17:45:43.12 &  -28:55:37.10  & 11.59 & 9.09 & 3600 & 3420 & 0.74 & $+$0.10   & 8.0  & $-$304.9  & 70        \\
GC37  & 17:45:35.94 &  -28:58:01.43  & 11.50 & 9.00  & 3600 & 3754 & 0.93 & $-$0.04   & 7.7  & $-$104.0  & 70      \\
GC44  & 17:45:35.95 &  -28:57:41.52  & 11.58 & 9.24  & 3600 & 3465 & 0.83 & $+$0.29   & 6.9  & $-$136.3  & 70      \\
\hline
\multicolumn{4}{l}{Disk stars}\\
\hline
$\alpha$ Boo$^b$ &  	14:15:39.67 & $+$19:10:56.7 & \multicolumn{2}{c}{$-$2.91} & $\mathrm{atlas}$ & 4286 & 1.66 & $-$0.52    & 6.3 & $+$26.5 &  250 \\ 
BD-012971     & 14:38:48.04 & $-$02:17:11.5 & \multicolumn{2}{c}{4.30} & 16 & 3573 & 0.50 &  $-$0.78    & 9.3 & $+$21.5 & 150 \\
142173$^{c}$  & 00:32:12.56 & $-$38:34:02.3 & \multicolumn{2}{c}{9.29} & 240 & 4330 & 1.50 &  $-$0.77   & 9.3 & $-$86.0 & 130\\
171877$^{c}$  & 00:39:20.23 & $-$31:31:35.5 & \multicolumn{2}{c}{8.12} & 240 & 3975 & 1.10 &  $-$0.93   & 7.0 & $-$30.5 & 150 \\ 
225245$^{c}$  & 00:54:46.38 & $-$27:35:30.4 & \multicolumn{2}{c}{8.79} & 240 & 4031 & 0.65 &  $-$1.16   & 6.6 & $-$48.3 & 130 \\ 
313132$^{c}$  & 01:20:20.66 & $-$34:09:54.1 & \multicolumn{2}{c}{7.04} & 120 & 4530 & 2.00 &  $-$0.20   & 8.7 & $-$54.4 & 190  \\ 
343555$^{c}$  & 01:29:42.01 & $-$30:15:46.4 & \multicolumn{2}{c}{8.46} & 240 & 4530 & 2.25 &  $-$0.74   & 6.8 & $-$161.7 & 240  \\ 
HD 787        & 00:12:09.99 & $-$17:56:17.8 & \multicolumn{2}{c}{1.86} & 4 & 4020 & 0.85 &  $-$0.16     & 8.2 & $-$39.7 & 110\\
\hline
\end{tabular}
\tablefoot{\tablefootmark{a}{K magnitude of the GC stars from Nishiyama (private comm.)
 and of the disk stars from 2MASS.}\\
\tablefootmark{b}{The fundamental parameters of $\alpha$ Boo are taken from \citet{aboo:param}.}\\
\tablefootmark{c}{Identification number from the Southern Proper-Motion Program \citep[SPM III][]{girard:04}, as given in \citet{monaco:11}. \\}}
\end{table*}

\begin{figure}
  \centering
	\includegraphics[width=0.5\textwidth]{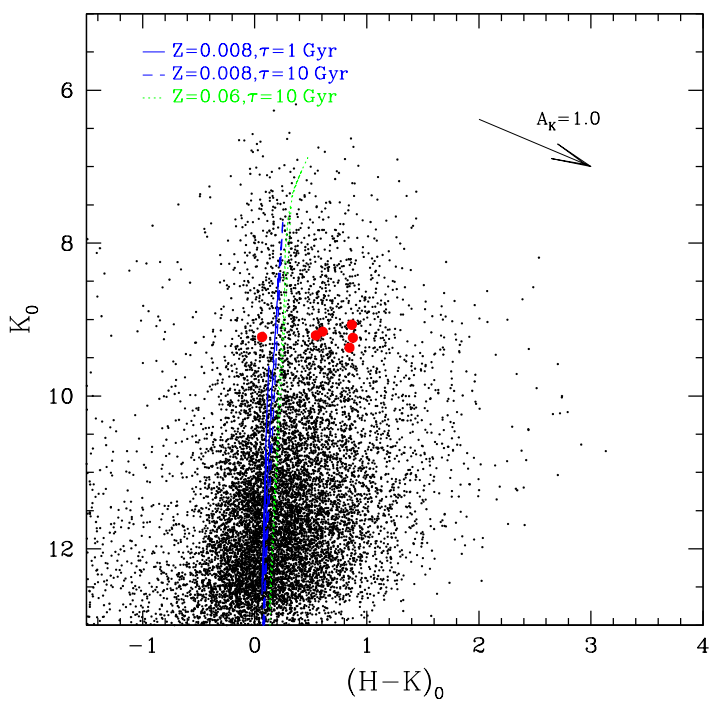}
	\caption{Dereddened $(H-K)_{0}$ vs. $K_{0}$ diagram of the GC field using the data if Nishiyama et al and the extinction map
of Schultheis et al. (2009).  In red are indicated our selected M giants. The blue lines show the Padova isochrones with Z=0.008 and $\rm \tau = 1\,Gyr$ (solid line) and $\rm \tau = 10\,Gyr$ (dashed line) while  the green line at Z=0.06 and $\rm \tau = 10\,Gyr$. The arrow on the top right shows the extinction vector of $\rm A_{K} = 1\,mag$ (Nishiyama et al. 2009)} 
	\label{cmd}
\end{figure}

\section{Observations} \label{observations}

When selecting our GC stars, we used the \citet{nishiyama2009} dataset and dereddened our sources using the
extinction map of \citet{schultheis2009}. Due to the high extinction, we
selected  field red giants (avoiding cluster and supergiant stars)  based on the dereddened $(H-K)_{0}$ vs. $K_{0}$ diagram covering the full $(H-K)_{0}$ colour range to avoid any selection bias.  Figure \ref{cmd} shows the dereddend  CMD together with our selected M giants  superimposed by the Padova isochrones with Z=0.008 and Z=0.06 and ages of 1\,Gyr and 10\,Gyr. Unfortunately the H--K colour
is very insensitive to the age of  population making it impossible to constrain the age and therefore the mass of our stars. 
  In addition,
to ensure that
our sources are located  in the GC itself, we used the high-resolution, 3D interstellar-extinction maps of the galactic Bulge of \citet{schultheis:14}, 
who used the VVV data set in distance bins of 500 pc. We placed our stars on these 3D extinction plots and found that our stars lie between $7-9$\,kpc which rules out any possible foreground contamination. We are therefore confident that all our `Bulge' stars, presented in in Table \ref{starsall}, belong to the GC.  The six local thick-disk stars, the thin-disk star BD-012971, and the reference star $\alpha$ Boo, which we also analyse in the same way as for the GC stars,  are also given in Table \ref{starsall}.

\begin{figure}
  \centering
	\includegraphics[width=\hsize]{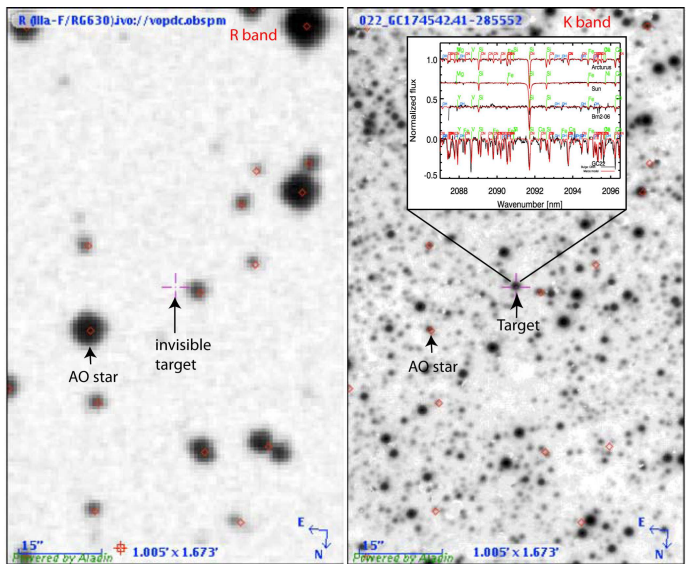}
	\caption{An example of a finding chart from our observing run with CRIRES at the VLT. The target star, GC22, shown here, is one of our 9 M giants in the Galactic Centre field, with stars only $2.5-5.5\,\arcmin$ North of the very center of the Milky Way. The left panel is an image in the R band at 660 nm (from the VO-Paris Data Centre in Paris, observed with the ESO Schmidt telescope), while the right one  shows the very same field but at $2.2$\,\mic\ \citep[a UKIDSS K-band image from DR9, see][]{ukidss}. The large difference in appearance is caused by the massive dust extinction  towards this field, 
prohibiting optical spectroscopy. The inset shows our K band spectrum of GC22, from which we can derive the Fe, Mg, Si, and Ca abundances. The {\it Adaptive Optics} (AO) system works in the R band.} 
	\label{RK}
\end{figure}

On 27-29 June 2012, we observed the M giants  close to the Galactic Centre and the disk stars given in  Table \ref{starsall}  in the near-infrared with the {\it Very Large Telescope, VLT}. Every star was observed both using the  low-resolution spectrometer ISAAC \citep{isaac}
and the high-resolution spectrometer CRIRES \citep{crires}, mounted on UT3 and UT1, respectively,  of the VLT. The ISAAC spectrometer, with a spectral resolution of  $R\sim1000$ and a wavelength range between $2.00-2.53$\,\mic, was used to determine the effective temperature of our stars, for details see Schultheis \& Ryde, in prep.
CRIRES uses, following standard procedures \citep{crires:manual}, nodding on the slit and jittering to reduce the sky background and the Adaptive Optics (AO) MACAO system to enhance the amount of light captured by the slit. Due to the large crowding in our field (see Figure \ref{RK}), special care had to be taken of the orientation of the $40''$ long slit. 
For the CRIRES observations, we used a $0.''4$ wide slit ($R\sim50000$), and  a standard setting ($\lambda^\mathrm{ref}_\mathrm{vac}=2105.5$, order=27) with an unvignetted spectral range covering $20818-21444$\,\AA, with three gaps (20\,\AA) between the four detector arrays. We aimed at a signal-to-noise ratio per pixel of 50-100 depending on the brightness of the stars. 
When selecting our stars, an instrumental constraint was set by the requirements that the wavefront sensing is done in the R band. For optimal AO performance, the guide star has preferably to be brighter than $R=14$ and  within $15''$ of the science object.
The reductions of the CRIRES observations were done by following standard recipes \citep{crires:cook} using Gasgano \citep{gasgano}. 
Subsequently, we used IRAF \citep{IRAF} to normalise the continuum, eliminate obvious cosmic hits, and correct for telluric lines (with telluric standard stars).   Two examples of CRIRES spectra are shown in Figure \ref{spectra}.






\section{Analysis}\label{analysis}

The derived stellar parameters of our stars are given in Table \ref{starsall}. \citet{ramirez1997}, \citet{ivanov2004}, and  \citet{blum2003}  studied  the behaviour of the $^{12}$CO band-head situated at 2.3\,$\mu$m with low-resolution K-band spectra and
found a remarkably  tight relation between the equivalent width and the effective temperature for M giants. We therefore analysed our ISAAC spectra guided by their work.
The analysis of our observations of additional calibration stars with known
effective temperatures, leads to  a tight relation between $\rm T_{eff}$ and the CO band intensity for $\rm  3200 < T_{eff} < 4000$ (Schultheis \& Ryde in prep.). The typical r.m.s  error  is in the order of 100\,K. We were able to apply this method for all but three of our stars; For the three warmer thick-disk stars, which have temperatures beyond the bound of our temperature calibration (the reason being that the CO-band strength weakens and is virtually unmeasurable), we had to resort to the temperature determination of \citet{monaco:11}, which also was  the source of our thick disk stars. 
These authors determined the effective temperatures by imposing a common Fe abundance from Fe\,{\sc i}  lines of different excitation energies.

The surface gravities of our GC stars were determined in a similar way as \citet{zoccali2008}  by adopting a mean distance of 8\,kpc to the GC, $T_{\odot}$ = 5770 K, $\log g_{\odot}$ = 4.44, $M_{\rm bol, \odot}$ = 4.75 and $M_\star = 1.0 M_{\odot}$. H and Ks band photometry from the VVV survey (\citealt{minniti2010}), extinction values from \citet{schultheis2009}, and the bolometric corrections from \citet{houdashelt2000} have been used.  
The surface gravity of all our disk stars are taken from the \citet{monaco:11} determination, except the thin disk star BD-012971, whose surface gravity was taken from \citet{origlia_M}.\\

Our CRIRES spectra were analysed using  the software  {\it Spectroscopy Made Easy, SME} \citep{sme,sme_code}  with a grid of MARCS spherical-symmetric, LTE model-atmospheres \citep{marcs:08}. SME is a spectral synthesis program that uses a grid of model atmospheres in which it interpolates for a given set of fundamental parameters. The program then derives an abundance from an observed spectral line by iteratively synthesising spectra which are compared to the observations by calculating the $\chi^2$. The program then finds the best fit, with the abundance as a free parameter, by minimising the $\chi^2$. For details, see \citet{sme}

An atomic line-list based on an extraction from the VALD database \citep{vald} is used. Due to inaccurate atomic data, atomic lines  were fitted to the solar intensity spectrum of \citet{solar_IR_atlas}, by our determining astrophysical $\log gf$ values for, most importantly,  Fe, and Si lines. The metallicity and Si abundance  we thus derive for $\alpha$ Boo are within 0.05 dex of the values determined by \citet{aboo:param}. However, since the Ca lines are very weak in the solar spectrum and the Mg lines are impossible to  fit, we have determined their line strengths such that they fit the Arcturus flux spectrum of \cite{arcturusatlas_II} instead, with a Mg and Ca abundance from \citet{aboo:param}; [Mg/Fe]$=0.37$ and [Ca/Fe]$=0.11$.  In the  syntheses we have also included a line list of the only molecule affecting this wavelength range, namely CN (J\o rgensen \& Larsson, 1990)\nocite{jorg_CN}.

The microturbulence,  $\chi_\mathrm{micro}$, is difficult to determine empirically, but is nevertheless important for 
the spectral syntheses, affecting saturated lines. We have chosen to use a typical value found in detailed investigations of red giant spectra in the near-IR by \citet{tsuji:08}. The microturbulence is more than a  km\,s$^{-1}$ larger  when determined form near-IR spectra compared to optical ones.
We have therefore chosen a generic value of $\chi_\mathrm{micro}=2.0$ km\,s$^{-1}$, see also the discussion in \citet{cunha2007}.  Furthermore, in order to match our synthetic spectra with the observed ones, we also convolve the synthetic spectra with a `macroturbulent' broadening, $\chi_\mathrm{macro}$, which takes into account the macroturbulence of the stellar atmosphere and instrumental broadening. These were determined by matching the prominent Si line at 20917.1\,\AA\ and the two Al lines at 21093.08\,\AA\ and 21163.82\,\AA.


\begin{figure*}
  \centering
	\includegraphics[width=\textwidth]{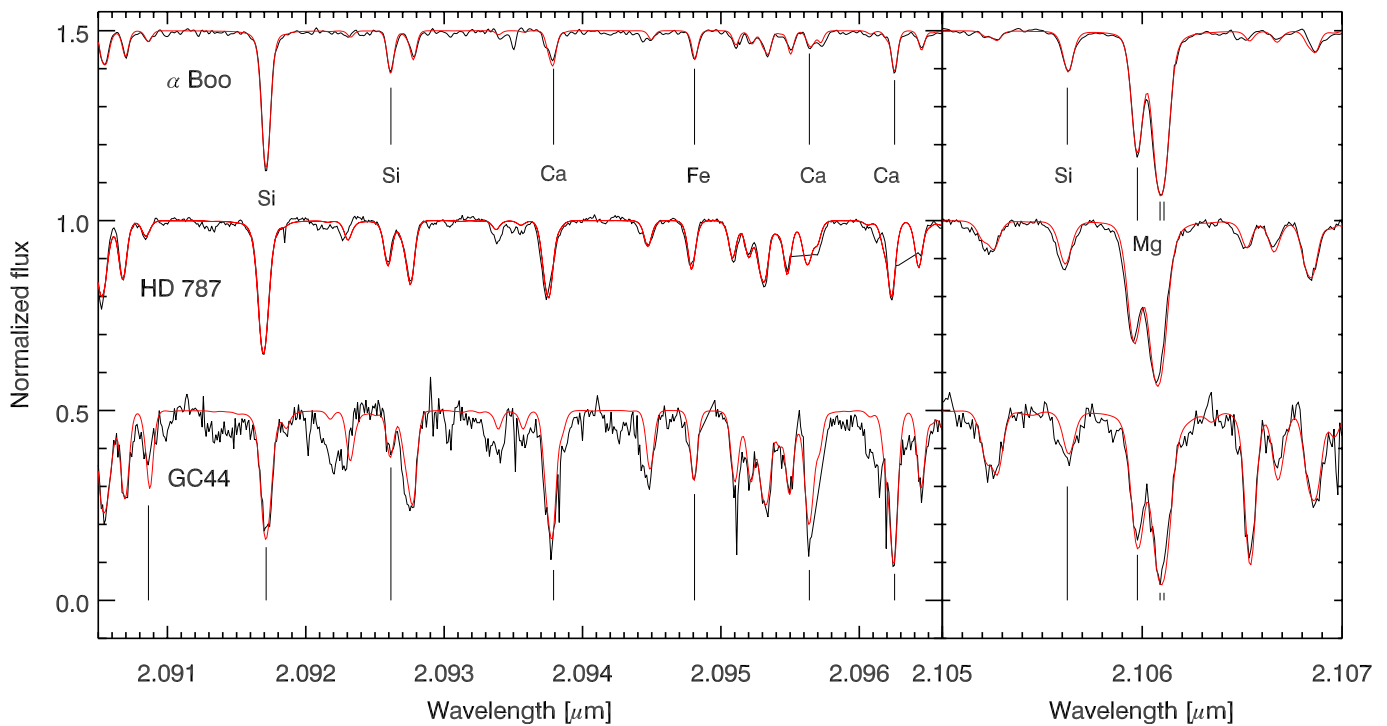}
	\caption{Spectra of wavelength regions covering a few of the lines used for the abundance determination. Other features not marked are mostly due to CN. The top spectrum is from the Arcturus atlas \citep{arcturusatlas_II}, the middle one is of the thick disk star HD 787, and the bottom spectrum is of the Galactic Centre star GC44. The flat regions of the HD 787 spectrum at 2.0955 and 2.0963\,\mic\ are due to the elimination of cosmic hits.} 
	\label{spectra}
\end{figure*}

\begin{figure*}
  \centering
	\includegraphics[width=\textwidth]{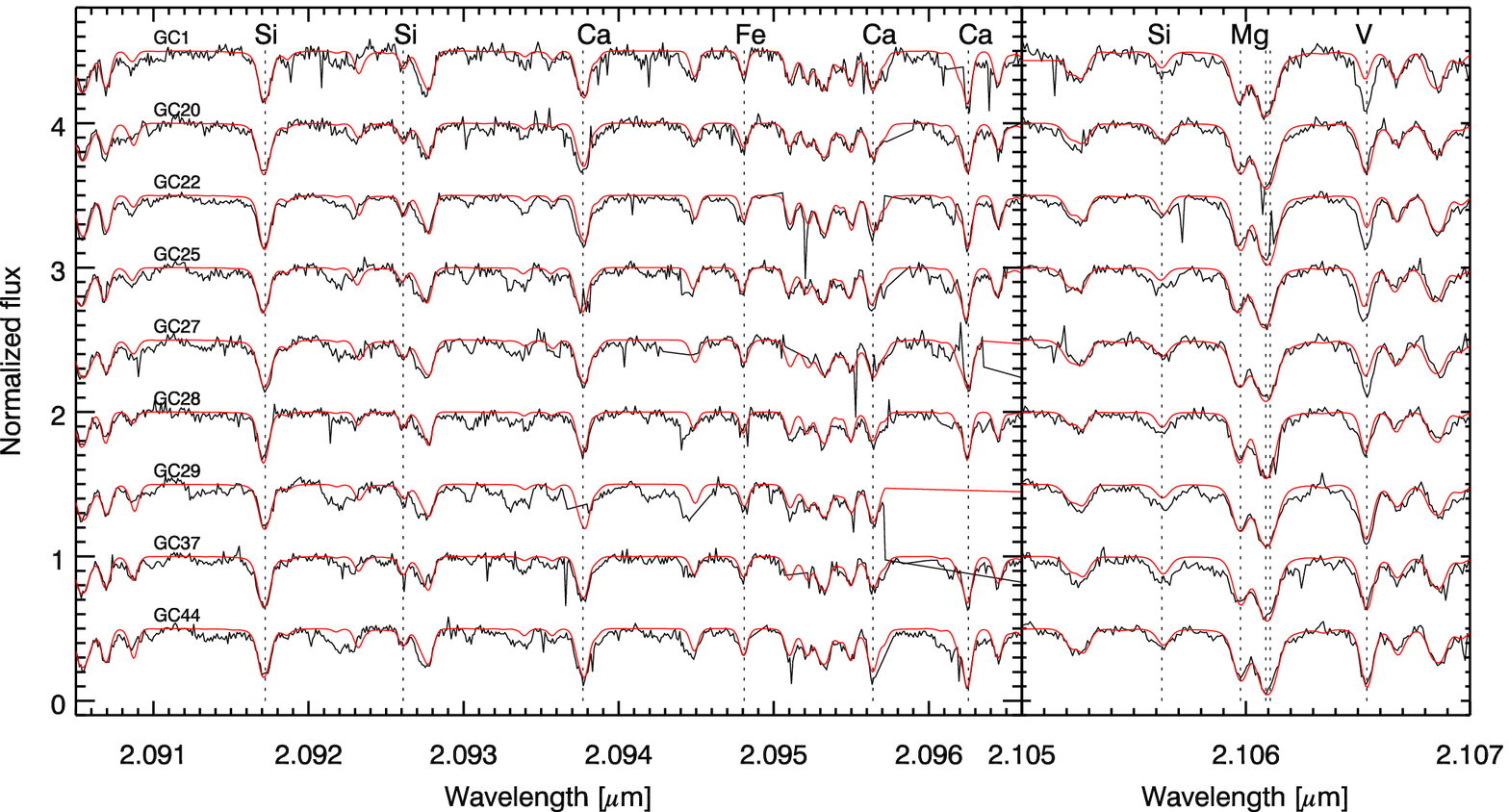}
	\caption{Observed spectra of all our Galactic Centre stars are shown in black. The best synthetic spectra are shown in red.} 
	\label{GC_spectra}
\end{figure*}

\begin{figure*}
  \centering
	\includegraphics[width=\textwidth]{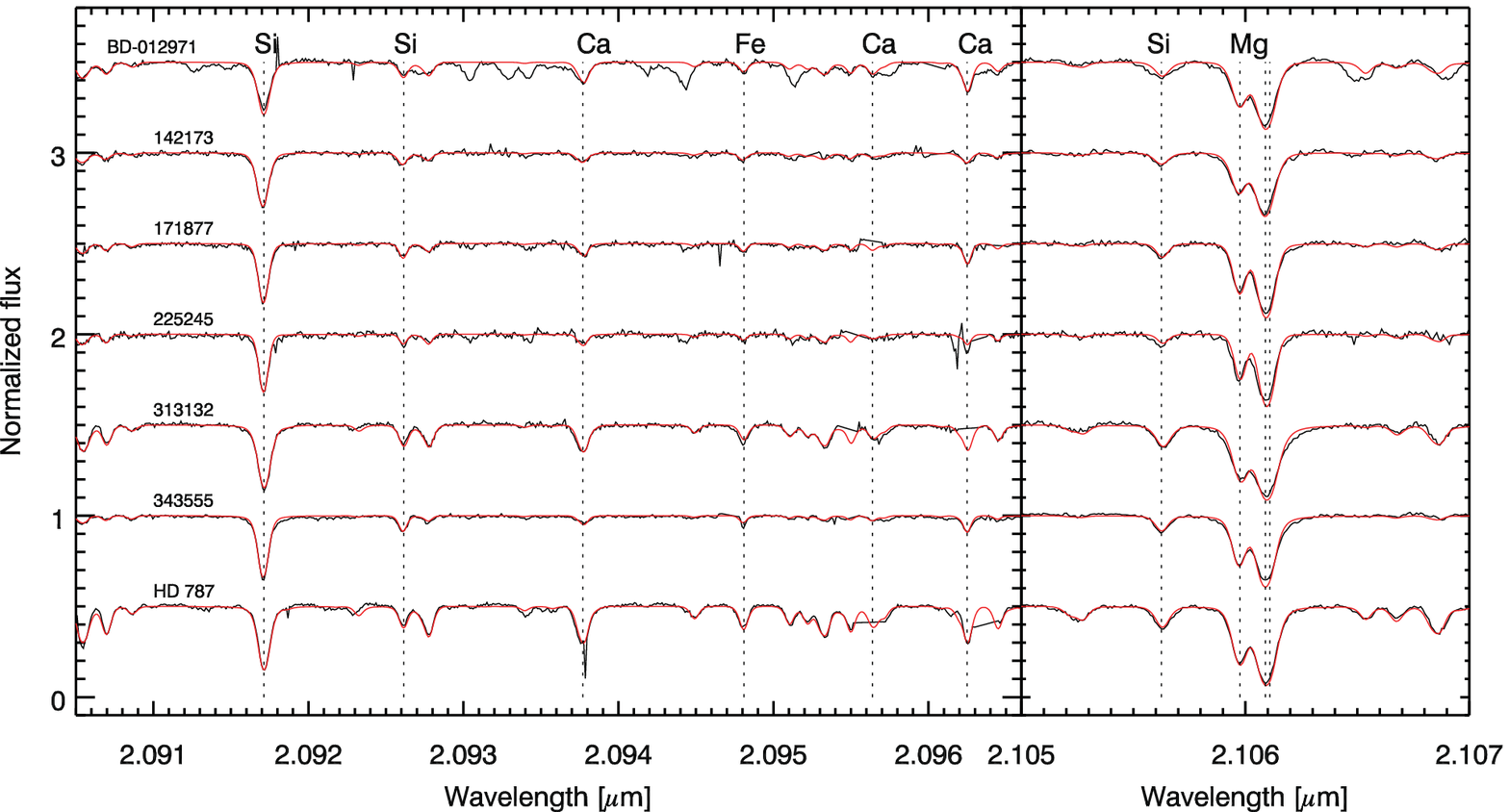}
	\caption{Observed spectra of all our Galactic Centre stars are shown in black. The best synthetic spectra are shown in red.  } 
	\label{disk_spectra}
\end{figure*}

The iron (giving the metallicity), Mg, Si, and Ca abundances were then determined by $\chi^2$ minimisation between the observed and  synthetic spectra for, depending on star, $6-9$ Fe, $2-6$ Si, $1-4$ Ca, $2$ Mg, and typically $10$ CN lines.
These abundances are presented in Table \ref{abund} and the synthesised spectra are shown with the observed ones for a few typical stars in Figure \ref{spectra}  and for all our spectra  in the online only Figures \ref{GC_spectra} and \ref{disk_spectra}. 

\begin{figure}
  \centering
	\includegraphics[width=0.5\textwidth]{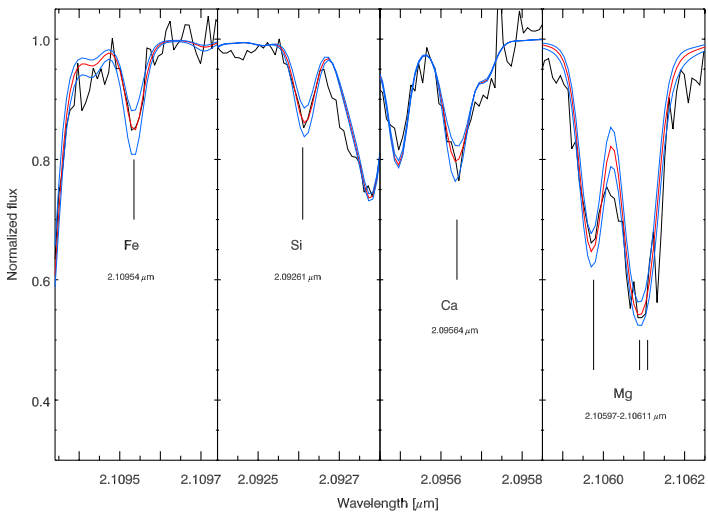}
	\caption{Sensitivity of typical spectral lines for a change of the abundance of Fe, Si, Ca, and Mg, respectively.  The red spectrum shows the best fit and the blue ones show the spectra for a change of $\pm 0.2 \,\mathrm{dex}$ in abundance. The star is in this case is GC28.}
	\label{pm}
\end{figure}

The uncertainties in the determination of the abundance ratios, for typical uncertainties in the stellar parameters are all modest, within 0.1 dex (see Schultheis \& Ryde, in prep.), see also Table \ref{errors}. Note, specifically the relatively small uncertainty in the abundance ratio due to a variation of the microturbulence parameter, $\xi_\mathrm{micro}$. This demonstrates that 
the  lines are not badly saturated, and therefore they should be sensitive to the abundances. 
In Figure \ref{pm} the sensitivity of the spectral lines to a change of $0.2\,\mathrm{dex}$ in abundance is shown. The lines are obviously useful for an abundance determination for these stars. 








\begin{table}
\caption{Derived abundances\label{abund}}
\centering
\begin{tabular}{lcccc}
\hline  \hline
Star & [Fe/H] & [Mg/Fe] & [Si/Fe] & [Ca/Fe] \\ 
\hline
GC1  &  0.19 & 0.11  & 0.12  & 0.16  \\
GC20 &  0.18 & 0.10  & 0.10  & 0.21  	\\
GC22 &  0.02 & 0.19  &  0.12 & 0.11  \\
GC25 &  0.16 & $-0.10$ & $-0.15$ &  0.03 \\
GC27 &  0.31 & 0.03  & 0.18  & $-$0.04 \\
GC28 & $-$0.10  & 0.14  & 0.05  & 0.27  \\
GC29 &  0.10 & 0.09  & 0.00  & 0.26  \\
GC37 & $-$0.04 & 0.13  & 0.18  & 0.30  \\
GC44 &  0.29 & 0.03  & $-$0.12 & 0.12  \\
\hline
$\alpha$ Boo & $-$0.53 & 0.37$^a$ & 0.33  & 0.11$^a$ \\
BD-012971  &  $-$0.78  &  0.32 & 0.29 & 0.28  \\
HD 142173  &  $-$0.77  &  0.28  & 0.28  & 0.16   \\
HD 171877  &  $-$0.93  &  0.47  & 0.46  & 0.38   \\
HD 225245  &  $-$1.16  &  0.46  & 0.36  & 0.26     \\
HD 313132  &  $-$0.20  &  0.32  & 0.30 & 0.27   \\
HD 343555  &  $-$0.74  &  0.45  & 0.45  & 0.42   \\
HD 787     &  $-$0.16  &  0.15  & 0.09   & 0.15    \\
\hline
\end{tabular}
\tablefoot{\tablefootmark{a}{Mg and Ca abundances from \cite{aboo:param}.}}
\end{table}

\begin{table}
\caption{Uncertainties in the derived abundances due to typical variations (or uncertainties) in the stellar parameters for a typical star with $T_\mathrm{eff}=3700$~K, $\log g = 1.5$, $\xi_\mathrm{micro}=2.0$~km\,s$^{-1}$, and solar metallicity. \label{errors}}
\centering
\begin{tabular}{lcccc}
\hline  \hline
Uncertainty & $\delta \mathrm{[Fe/H]}$ & $\delta \mathrm{[Mg/Fe]}$ & $\delta \mathrm{[Si/Fe]}$ & $\delta \mathrm{[Ca/Fe]}$ \\
\hline
$\delta T_\mathrm{eff} = +150\,\mathrm{K}$ & $+$0.02 & $-$0.02 & $-$0.10 & $+$0.07\\
$\delta \log g = +0.5$ & $+$0.02 & $-$0.02 & $\pm0.00$ & $-0.01$ \\
$\delta \mathrm{[Fe/H]} = +0.1$ &   & $-$0.08 & $-$0.09 & $-$0.08 \\
$\delta \xi_\mathrm{micro} = +0.5$ & $-$0.10 &  $-$0.07 & $-$0.03 & $+$0.06\\
\hline
\end{tabular}
\end{table}

\begin{figure*}
  \centering
	\includegraphics[width=\textwidth]{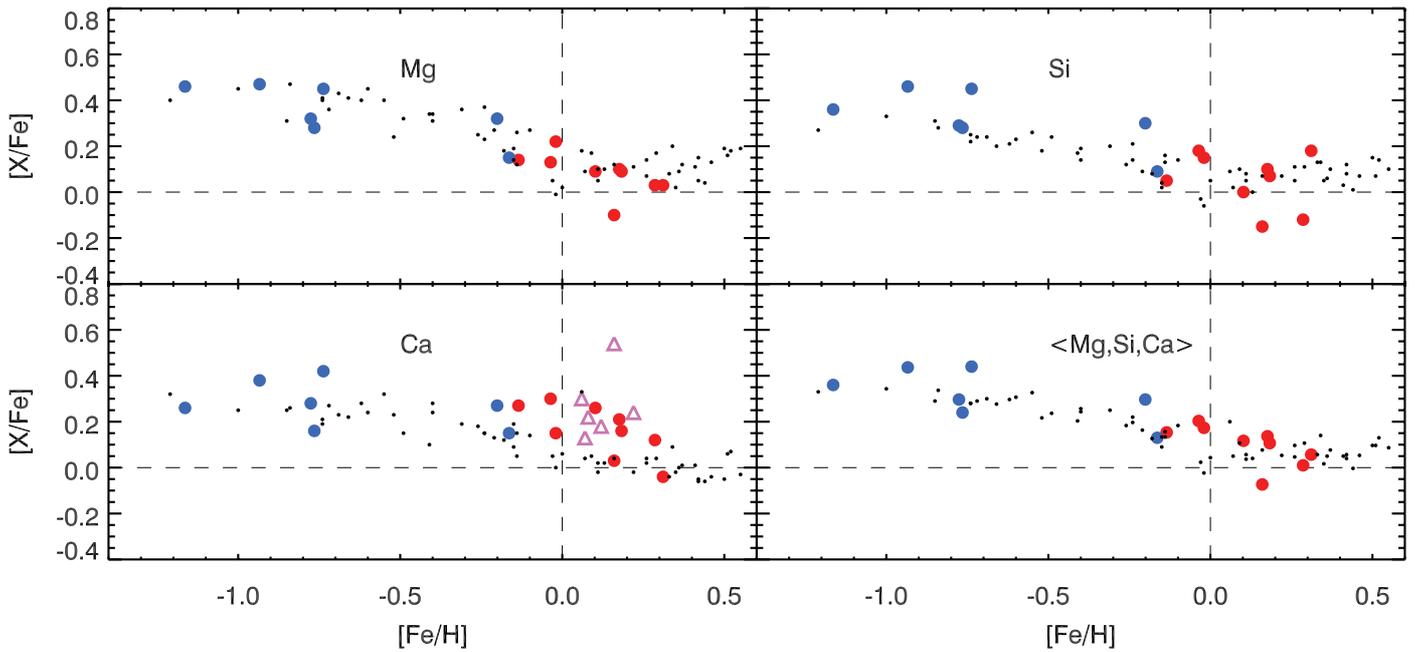}
	\caption{Abundances trends of [Mg/Fe], [Si/Fe], and [Ca/Fe], as well as the mean of these. Red dots mark our Galactic Centre stars, blue dots our local disk star measurements, and magenta triangles mark the [Ca/Fe] measurements (with uncertainties of $\pm0.15$ dex) of M giants (with $M_\mathrm{bol}>-5.5$) in the GC by \citet{cunha2007}. The black small dots are the abundances determinations based on micro-lensed dwarfs by \citet{bensby:13}. } 
	\label{afe}
\end{figure*}

\begin{figure}
  \centering
	\includegraphics[width=0.5\textwidth]{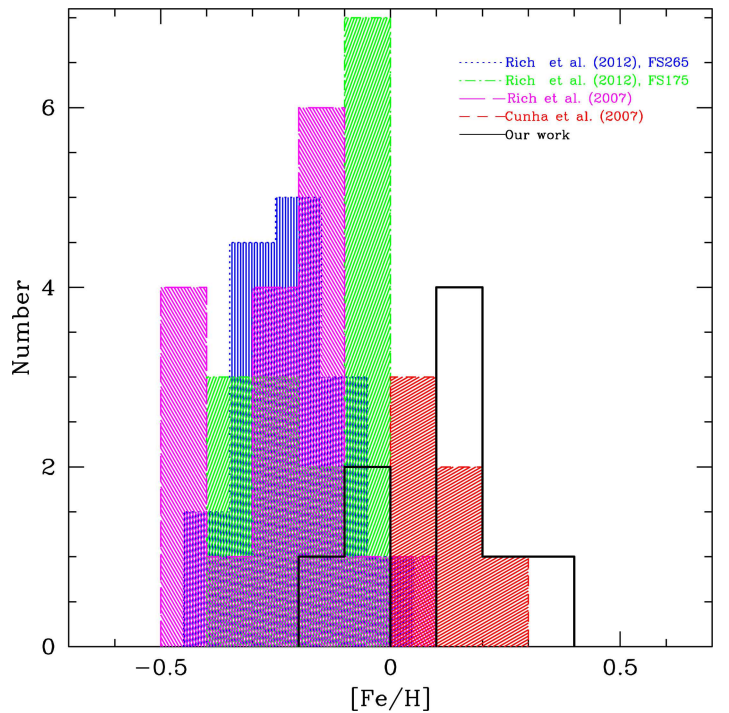}
	\caption{Histogram of our metallicities in black with a mean of [Fe/H]=0.11 and a standard deviation of 0.15. The red histogram shows the metallicities of the Galactic centre giants of \citet{cunha2007}. For comparison we  show in magenta the field of  Rich et al. (2007) located at ($l$,$b$)$=$($0^{\circ}$,$-1^{\circ}$), in green the field at ($0^{\circ},-1.75^{\circ})$ (Rich et al. 2012) and in blue the field at ($1^{\circ},-2.65^{\circ})$ (Rich et al. 2012).}
	\label{hist}
\end{figure}

\section{Results and Discussion} \label{Discussion}

Figure \ref{afe} shows the abundance ratio trends for Mg, Si, and Ca as derived from our GC stars in red and our local thick-disk stars in blue. In black we show the micro-lensed bulge dwarfs from the outer bulge ($|b|>-2^\circ$) as determined by \cite{bensby:13}. Within uncertainties, our thick disk trend as well as the Mg and Si trends follow the outer bulge trends, with low values of [$\alpha$/Fe] at [Fe/H]$>-0.2$, although the GC stars show a much more narrow spread in [Fe/H]. Our [Ca/Fe] in the GC show a higher trend than the outer bulge stars, actually in agreement with \citet{cunha2007} and one of the outer bulge stars. With random uncertainties being of the order of $0.1$ dex, this trend is significantly higher. This difference between Mg, Si, and Ca is not expected theoretically (Matteucci, private comm.), and systematic uncertainties could be important. In order to reduce random and systematic uncertainties, we therefore also plot a mean of the three trends in the lower right panel. From this plot it is evident that we can not claim that our [$\alpha$/Fe] trend in the GC is particular compared to the outer bulge. However, more investigations on the [Ca/Fe] trend is needed to verify its high trend in the GC. 

Figure \ref{hist} shows the tight metallicity distribution with [Fe/H]=$+0.11\pm0.15$ and the total spread in  metallicities of 0.41 dex. Our mean metallicity distribution is close to
that found by \citet{cunha2007}, indicated by the red histogram.  However, our total spread  is broader than theirs of $0.16$, but it is much smaller than found typical  for giants in the Galactic Bulge (see e.g. the metallicity distributions  of 
\citet{zoccali:2003}, \citet{fulbright:06}, 
\citet{hill2011}), and \citet{ness:13}.  
The most striking feature is clearly the absence of the metal-poor population with no stars below $-0.1$\,dex in [Fe/H] in agreement with the work of \citet{cunha2007}.  This strengthens the argument of a specific population different to those
of the   Bulge. Fields in the inner Bulge located   at $b=-1^{\circ}$ \citep{rich2007} (magenta) and $b=-1.65^{\circ}$ (green) as well as $b=-2.65^{\circ}$ (blue) \citep{rich:12}  have mean metallicities at [Fe/H]=$-0.22$,  [Fe/H]=$-0.16$, and [Fe/H]$=-0.21$, which are $\sim$ 0.3\,dex lower than our metallicities in the GC (see Fig. \ref{hist}).

\begin{figure*}
  \centering
	\includegraphics[width=\textwidth]{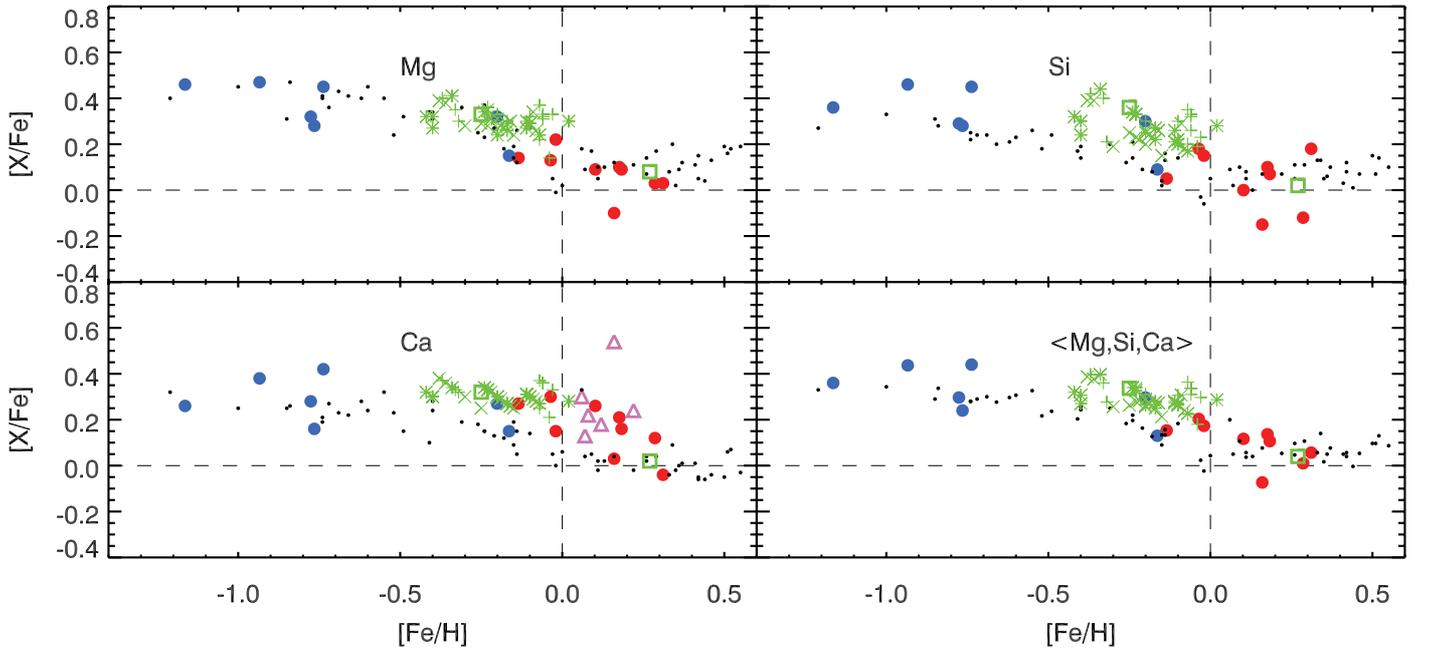}
	\caption{Same abundance ratio trends as in Figure \ref{afe}, but with the addition of the trends derived from the M giants in the fields located at ($l$,$b$)$=$($0^{\circ}$,$-1^{\circ}$) of \citet{rich2007} denoted by green asterisks, and at ($0^{\circ},-1.75^{\circ}$) and ($1^{\circ},-2.65^{\circ})$ of \citet{rich:12}, denoted by green pluses and crosses, respectively. Furthermore, the mean abundance ratios of the two populations of Terzan 5 at $(l,b)=(3.8^\circ$,$+1.7^\circ)$ from \citet{origlia:11} are shown with green squares. For the inner globular cluster Liller 1  at $(l,b)=(354.8^\circ$, $-0.2^\circ)$, \citet{origlia_GC4} find [Fe/H]$\sim-0.3$ and [$\alpha$/Fe]$\sim+0.3$], which fits nicely with the general trend of the Rich et al. (2007, 2012) field stars.  }
	\label{afe_rich}
\end{figure*}

 In Figure \ref{afe_rich} we show our abundance-ratio trends again, but now together with those of 
the inner Bulge fields, as derived by \citet{rich2007} and \citet{rich:12}, shifted to the solar scale of \citet{marcs_solar}, which we use in our analysis.  It is clear again that our metallicities are generally higher, with a small overlap.  Although there are certainly systematic uncertainties that could make a comparison difficult, it is worth discussing;  Compared to the large difference in the metallicities of the stars, the [$\alpha$/Fe] abundance ratios agree quite well, although our values tend to be of the order of 0.1 dex lower. It is interesting to note that the [Ca/Fe] versus [Fe/H] trend seem to show higher values
from all the inner fields compared to the micro-lensed dwarf sample.

In Figure \ref{afe_rich} we also plot the mean abundances of the two populations in the complex globular cluster Terzan 5 located in the inner Bulge, $(l,b)=(3.8^\circ$,$+1.7^\circ)$, as derived by \citet{origlia:11} (in green squares in Figure  \ref{afe_rich}). It is interesting that the metal-rich population at [Fe/H]$= +0.27$,  shows solar [$\alpha$/Fe], similar to our GC population at that metallicity. However, in contrast to Terzan 5,  we clearly miss  a distinct metal-poor population. \citet{origlia:11} suggest that there might be a common origin and evolution of field populations in the Bulge and Terzan 5, the latter perhaps being a relic building block of the Bulge.

Thus, we see  a metal-rich but low $\alpha$-enhanced population in the GC. This finding would rule out a scenario that SNIIe played the dominated role in the chemical enrichment of the gas  associated with a rapid formation. Again we find  a significant  difference compared to the high $\alpha$-enhanced values of \citet{rich2007,rich:12} indicating that the GC population is different. The metal-rich population of Terzan 5 shows similarities to our GC population.  

The presence of a double bar in our Galaxy which can be observed in  external galaxies \citep{laine:02,erwin:04}  is still questioned. \citet{alard:01}, \citet{nishiyama:05}, and \citet{gonzalez:11}  claim that there is an inner structure distinct to the large-scale Galactic bar, with a different orientation angle which could be associated with a  secondary, inner bar. On the other hand,  \citet{gerhard:12} can explain this inner structure by dynamical
instabilities from the disk without requiring a nuclear bar.  The metal-rich metallicities together with low $\alpha$-enhancement  that we derive for the GC stars in our study presented here, seems to indicate  a bar-like population (as seen by \citet{babu:10} for the main bar)  most likely related to the nuclear bar.

Detailed chemical evolution models for the central 200 pc of the GC, taking into account the increased star formation history, star formation efficiency,  various gas-infall  and gas outflow mechanisms, including the possible role of the nuclear bar are unfortunately missing. Detailed chemical/dynamical modelling of this extreme part of our Galaxy is needed for a proper interpretation of the observables but also  for Bulge formation models in general.


\begin{acknowledgements}
The anonymous referee is thanked for suggestions that improved the paper.
Nils Ryde is a Royal Swedish Academy of Sciences Research
Fellow supported by a grant from the Knut and Alice Wallenberg Foundation, 
and 
acknowledges support from the Swedish Research Council,
VR (project number 621-2008-4245), 
Funds from Kungl. Fysiografiska S\"allskapet i Lund. 
(Stiftelsen Walter Gyllenbergs fond and M\"arta och Erik Holmbergs donation). 
This publication makes use of data products from the Two Micron All Sky Survey, which is a joint project of the University of Massachusetts and the Infrared Processing and Analysis Center/California Institute of Technology, funded by the National Aeronautics and Space Administration and the National Science Foundation.
This work has made use of the VALD database, operated at Uppsala University, the Institute of Astronomy RAS in Moscow, and the University of Vienna.
\end{acknowledgements}

\bibliographystyle{aa}

\Online

\end{document}